\documentclass[9pt,twocolumn,twoside]{osajnl}

\journal{jocn}

\setboolean{shortarticle}{false}
\usepackage{graphicx}
\usepackage{caption}
\usepackage{subcaption}
\usepackage{placeins}
\usepackage{lipsum}

\title{Estimation of von Neumann entropy of different light sources using array detectors}

\author[1]{R. P. George}
\author[1]{Tomis}
\author[1,2,*]{V. Narayanan}
\author[1,2]{Subhashish Banerjee}

\affil[1]{Departement of Physics, Indian Institue of Technology Jodhpur, Rajasthan,342037, India }
\affil[2]{Interdisciplinary Research Platform- Quantum Information and Computation(IDRP-QIC), Indian Institue of Technology Jodhpur, Rajasthan,342037, India }

\affil[*]{Corresponding author: vnara@iitj.ac.in}

\dates{}

\doi{}

\begin{abstract}
 We employ a method involving an array detector to measure the transverse spatial variation of the von Neumann Entropy (VNE) associated with the polarization state of light for different light sources including the coherent light from a diode laser, the chaotic light from an LED, the fluorescent emission from a dye widely used as a contrast agent and also the downconverted output from the process of Spontaneous Parametric Down Conversion which has widespread applications in Quantum Optics as a source of entangled photon pairs. Additionally, we studied the variation of the Stokes parameters as well as von Neumann entropy with pixel binning in the array detector output.
\end{abstract}

\setboolean{displaycopyright}{false}

\begin{document}

\maketitle

\section{Introduction}
The von Neumann Entropy[1] associated with the polarization of light has several interpretations; in accordance with the thermodynamic definition for entropy, it can be considered as a measure of randomness or from an information standpoint, it can be seen as the difference in the amount of information [2] between pure and mixed states of the same intensity. In the context of polarization, it exhibits an inverse relationship with the degree of polarization [1,3] , which is an expression of polarimetric purity. The von Neumann or polarization entropy can be thought of as another measure of polarimetric purity with its value tending to zero when the polarization state is pure and approaching unity when it is completely mixed. 

The von Neumann entropy is derived from the normalized coherency matrix [1] which can uniquely characterize any polarization state. This connects it with the concept of Stokes polarimetry [4] and opens up new possibilities in the field of polarimetric imaging which has been seeing a lot of interest lately.
In this work we have developed a simple yet scalable setup to measure the transverse spatial variation of the polarization von Neumann Entropy across the dimensions of the illuminated area of the array detector.

Additionally, we extracted the transverse spatial variations of all normalized Stokes parameters as well as the sum of their squares. We also studied the effects of pixel binning on VNE and square sum values.
The light from a diode laser at 405 nm, an LED, a fluorescent dye and the polarization entangled SPDC output from a BBO crystal phasematched for type-II downconversion using a 405 nm pump were studied using our setup. 

This has applications in several fields, primarily as a means of studying the variations in polarization brought about by a medium [5-14] during the propagation of polarized EM waves through said medium and also in Lidar imaging, remote sensing and astronomy [15]. Polarimetry based studies have been extensively investigated for the measurement of magnetic field in Laser produced plasmas [16].

\section {Stokes Parameters and polarization von Neumann entropy}
\label{sec:examples}

The concept of Stokes polarimetry has been vastly explored and reported in exceptional details [1, 3-6]. The four Stokes parameters are sufficient to fully describe any polarization state; pure or otherwise. They are derived from a set of projective measurements made on the unknown state of light. A set of six measurements are commonly adopted in polarimetric measurement [6] to yield the normalized Stokes parameters $s_{0},s_{1},s_{2}$ and $s_{3}$

\begin{equation}
\begin{aligned}
s_{0} &=\dfrac{I_{H}+I_{V}}{I_{H}+I_{V}}, && 
s_{1} =\dfrac{I_{H}-I_{V}}{I_{H}+I_{V}},\\
s_{2} &=\dfrac{I_{A}-I_{D}}{I_{A}+I_{D}}, &&
s_{3} =\dfrac{I_{R}-I_{L}}{I_{R}+I_{L}}.
\end{aligned}
\end{equation} 

Here $I_{V},I_{H},I_{A},I_{D},I_{R}$  and $I_{L}$ correspond to the six intensity measurements along vertical, horizontal, anti-diagonal, diagonal, right circular and left circular bases respectively. Using a combination of Stokes parameters and Pauli’s spin matrices [1], it is possible to represent the polarization state of light using a coherency matrix [1]

\begin{equation}
\Phi=\dfrac{1}{2}\sum_{i=0}^{3} s_{i}\sigma_{i}=\dfrac{1}{2}\begin{bmatrix}
s_{0}+s_{1} & s_{2}-is_{3}\\
s_{2}+is_{3} & s_{0}-s_{1} 
\end{bmatrix}.
\end{equation}

Here, $\sigma_{i}$ represent the four Pauli matrices and $\Phi$ represents the coherency matrix. A coherency matrix is a 2x2 matrix that is positive semi-definite, Hermetian and contains all measurable second order information about the 2D polarization state.

Once the coherency matrix is realized, it is possible to arrive at the von Neumann entropy associated with the polarization state using the following relation [1]
\begin{equation}
S_{VN}= -tr\{\Phi log_{2}\Phi\}.
\end{equation}
Alternatively, the von Neumann entropy can be arrived at from the normalized eigenvalues {of the coherency matrix $\Phi$, represented by} $\lambda_{1}$ and $\lambda_{2}$
\begin{equation}
S_{VN}= -\sum_{i=1}^{2}\{\lambda_{i} log_{2}\lambda_{i}\}.
\end{equation}
When the value of $S_{VN}$ approaches 0, the state is very nearly pure, i.e. is almost completely polarized and becomes mixed or unpolarised as $S_{VN}$ nears unity.

Alternatively, one can turn to the degree of polarization ($\mathcal{P}$) as another metric to quantify state purity from the Stokes parameters using [1,2,6]
\begin{equation}
\mathcal{P}= \dfrac{\sqrt{s_1^2+s_2^2+s_3^2}}{s_{0}}.
\end{equation}
While $\mathcal{P}$ describes a different attribute as compared to the von Neumann entropy, they are related by the expression [1]
\begin{equation}
S_{VN}=-[\dfrac{1+\mathcal{P}}{2} log_2 (\dfrac{1+\mathcal{P}}{2})+\dfrac{1-\mathcal{P}}{2} log_2 (\dfrac{1-\mathcal{P}}{2})].
\end{equation}
This shows a one to one correspondence between the von Neumann entropy and the degree of polarization, validating the use of $S_{VN}$ as a quantifier of polarimetric purity.

\section{Measuring the Stokes Parameters using an array detector}

Much like single pixel Stokes polarimetry, a combination of waveplates and linear optics need to be adopted for this purpose. The Electron Multiplied Charge Coupled device (EMCCD, Mode; PRO-1024BX:Excelon) was employed as an array detector in this investigation. A very weak incident photon signals are amplified greatly to a level above the inherent noise with the EMCCD and this feature opens up several applications for the device. Adopting the basic setup for polarimetry, it is possible to examine the spatial variation of polarization of a very weak optical signals using the EMCCD, as shown in Fig. 1. The intensity values from each detector pixel can be used to derive the Stokes Parameter values for the same. Furthermore knowing the Stokes parameters, we can derive the coherency matrix corresponding to every pixel and ultimately the von Neumann entropy corresponding to each pixel.

An efficient method would be to utilize the available detector area on the EMCCD by employing some sort of polarizing beam splitter like a Glan-Taylor polarizer or a Wollaston prism and splitting the beam into orthogonally polarized components that are allowed to fall on spatially separated regions of the Sensor array. Of course finding the Stokes parameters and all the associated quantities for each and every pixel will be computationally intensive and as a consequence, time consuming. We therefore will focus on the two regions of interest corresponding to the orthogonally polarized modes emergent from the polarizing beam splitter. Furthermore, the device allows for binning of intensity data from individual pixels and the effect of this binning on Polarization parameter values needed examination. The intensity values from each pixel is binned into groups with a user defined number of pixels in each group. The advantage of using the EMCCD is that the necessary information can be extracted post experiment and processed using a simple program written in Matlab as the EMCCD supports the export of experimental data with a communicating computer.

Our main setup relied on a Glan Taylor polarizing (GTP) beam splitter to separate the orthogonal components of the attenuated laser beam and while the EMCCD was aligned to image the direct propagating mode, the other mode was redirected onto the EMCCD sensor using a mirror. We also employed a setup with a Wollaston prism replacing the Glan Taylor polarizer, negating the need for alignment mirrors just before the EMCCD.

\begin{figure}[h]
\centering     
\includegraphics[width=100mm]{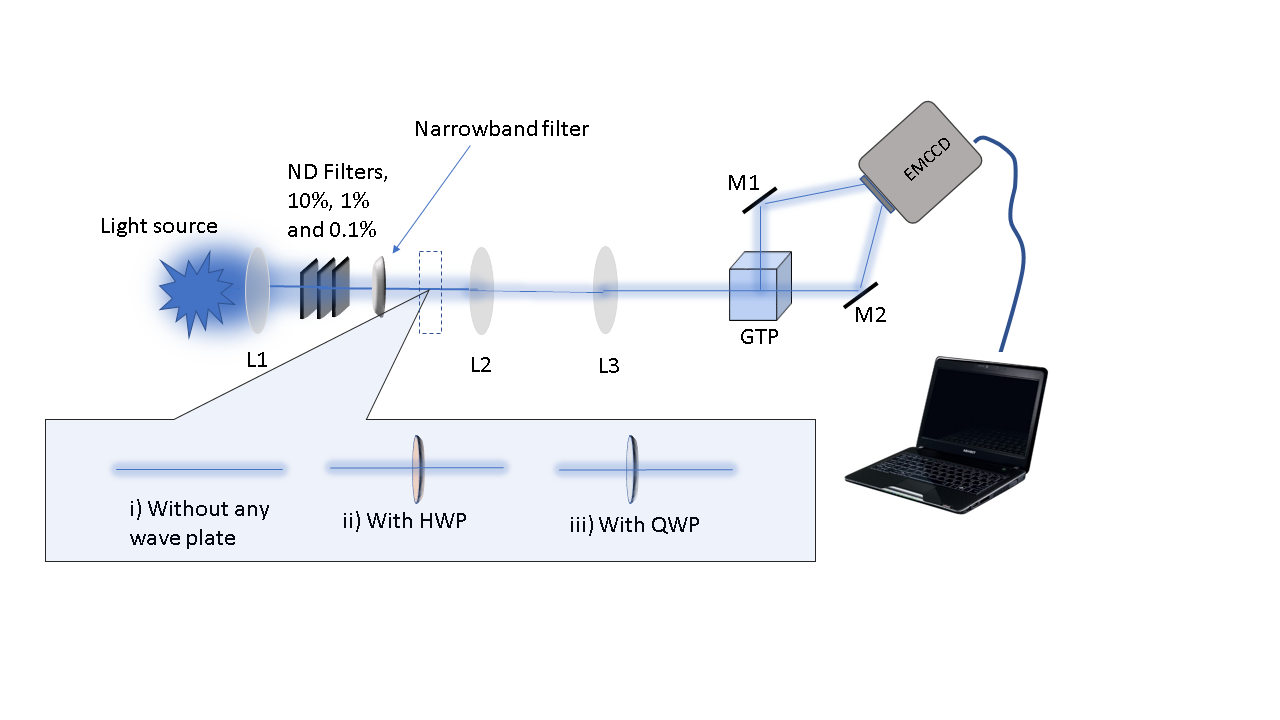}
\caption{The measurements are made in three steps with different waveplates.  {In the experimental setup, L1, L2 and L3 are lenses, ND refers to Neutral Density filter used to attenuate the light intensity, M1 and M2 are mirrors and GTP is a Glan Taylor Polarizer.}}
\label{fig:a}
\end{figure}

A total of three images are acquired using the EMCCD with the waveplates set to different settings corresponding to the required measurements, as shown in Fig. 1. A Matlab program was written to read the output files from the EMCCD and to process the output files, intensity profiles from the region of interest were recorded appropriately. The resulting data is a set of six intensity valued arrays, with each array corresponding to the intensity values recorded by each pixel within the region of interest for the corresponding measurement setting. According to the formulae to calculate the Stokes parameters the program then performs the required mathematical operations with the acquired arrays. Once the Stokes parameters corresponding to each pixel in the region of interest are obtained and stored, the program will move on to the process of using the freshly obtained Stokes parameters to compile the set of coherency matrices also corresponding to every pixel from the region of interest. Finally, with the coherency matrices generated, the program will derive the von Neumann entropy values corresponding to every pixel in the region of interest.

Also, we had decided to go with three different bin sizes (hardware bins) for the experiment and study the variation across these. The first involved no binning resulting in a 1024 by 1024 array of intensity values corresponding to the flux incident on each physical pixel. The second case involved binning the intensity of 4 pixels into one (2x2 binning), resulting in a 512 by 512 array of intensity values. The final case would give a 256 by 256 intensity array after binning 16 pixels into one (4x4 binning). Binning has the added effect of improving the signal to noise ratio at the expense of spatial resolution. This became apparent when the same acquisition settings that could be considered the bare minimum for a proper signal-to-noise ratio (SNR) in the case of no binning, caused the sensor to saturate in the 4x4 binned state. This forced us to adopt acquisition settings, in terms of exposure time and gain, such that the sensor will not saturate in the maximum binned state but at the cost of poor SNR ratio in the 1024 by 1024 unbinned state.
In the coming sections we will discuss our findings using the different light sources.





\section{Diode Laser}

The laser we used was a vertically polarized CW diode laser (iflex-2000, Qioptiq)  with output peaking at 405 nm and with about 2 nm of Full-width half maxima (FWHM). This source was used to calibrate the setup because of the very well defined polarization of its output. Our array detector, being an EMCCD, requires the use of a combination of several neutral density filters to attenuate the laser output down to an intensity that will not damage our EMCCD. A combination of GTP with mirrors was utilized to image the orthogonal components of the beam using EMCCD.

The effect of the reduced SNR shows in our calibration experiment where parameters like the s1 value, which should be close to negative unity but shows a spatial average of around half of that. However, for both binned cases, the values are very much in agreement with the theory and we attribute this to the improved SNR associated with the binning process. We had to compromise SNR in the case of the unbinned measurement to maintain uniform measurement parameters of exposure time and EMCCD gain  across all binning conditions. The device sensor quickly exceeds its safe limit when binning is applied if SNR of the unbinned measurement is kept high using large gain values or increased exposure times.
The colormaps shown in Fig. 2. depict the profile of von Neumann entropy for the diode laser and it is evident from the scale how minute the variations are in this case. It is also clear from Fig. 2. how binning reduces the resolution.
Fig. 3 shows how binning affects the Stokes parameter values  as well as von Neumann entropy and the square sum of the Stokes parameters. The plot exhibits the inverse proportionality of the squared sum of the
Stokes parameters to the von Neumann entropy for different binning conditions.

\begin{figure}[!hbt]
     \centering
     \begin{subfigure}[b]{0.9\columnwidth}
         \centering
         \includegraphics[width=\textwidth]{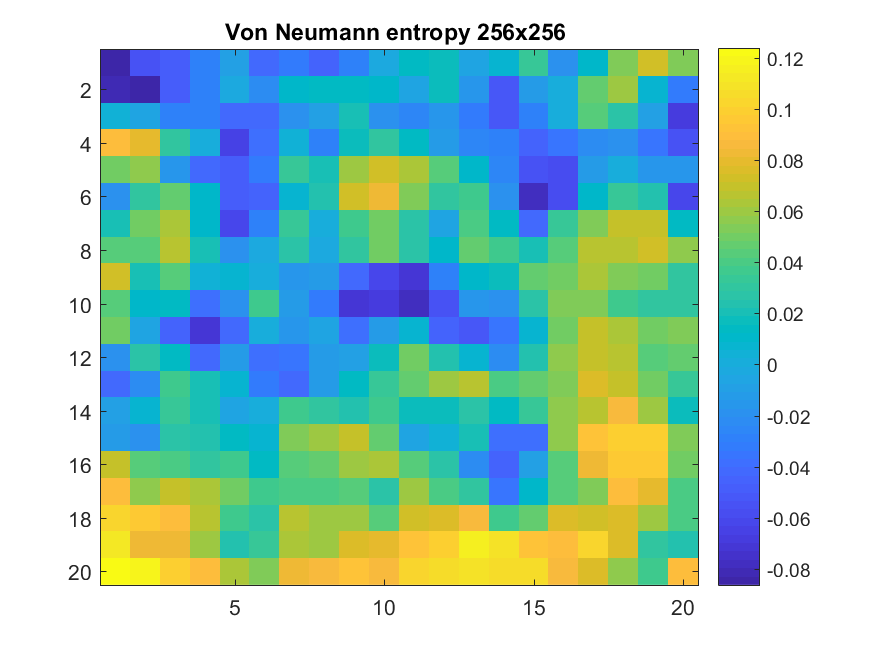}
         \caption{4$\times$4 binning}
         \label{fig:a}
     \end{subfigure}
     \hfill
     \begin{subfigure}[b]{0.9\columnwidth}
         \centering
         \includegraphics[width=\textwidth]{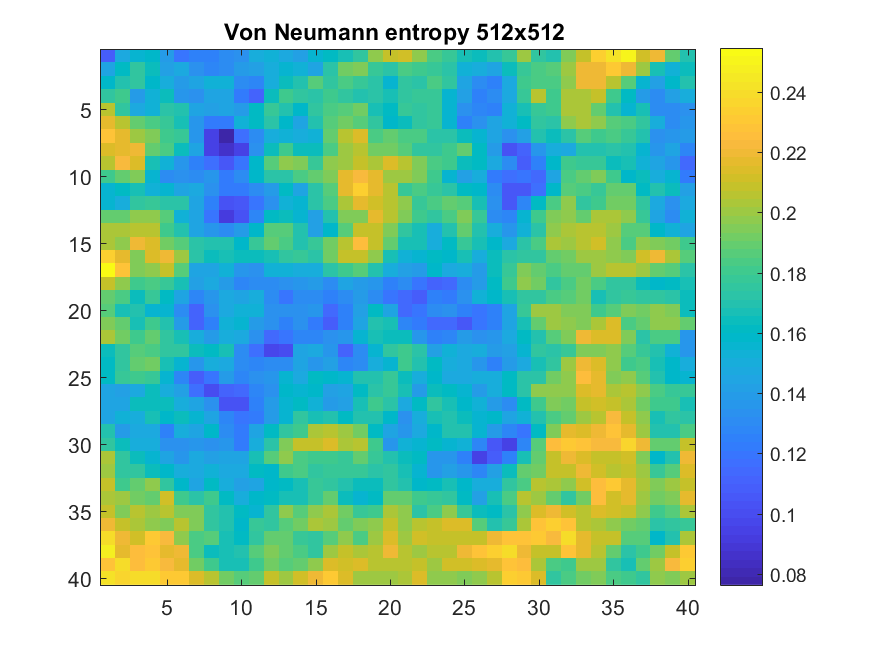}
         \caption{2$\times$2 binning}
         \label{fig:b}
     \end{subfigure}
     \hfill
     \begin{subfigure}[b]{0.9\columnwidth}
         \centering
         \includegraphics[width=\textwidth]{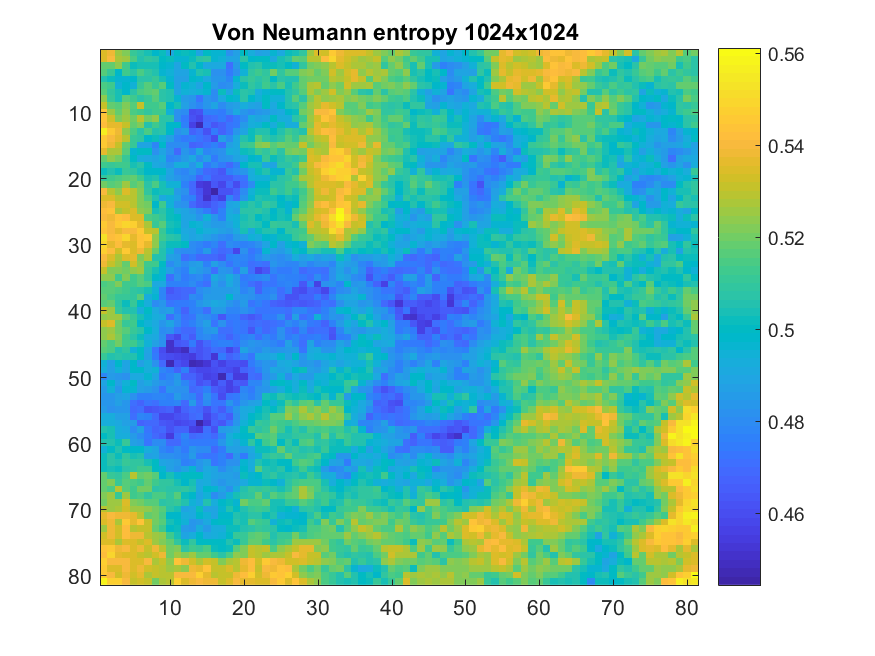}
         \caption{1$\times$1 binning}
         \label{fig:c}
     \end{subfigure}
        \caption{The spatial variation in the von Neumann entropy for different bin sizes as shown in each map title. These correspond to the case of light from the diode laser. The variations, although apparently drastic, are actually spread over a very narrow range.}
        \label{fig:2}
\end{figure}

\begin{figure}[!hbt]
     \centering
     \begin{subfigure}[b]{0.9\columnwidth}
         \centering
         \includegraphics[width=\textwidth]{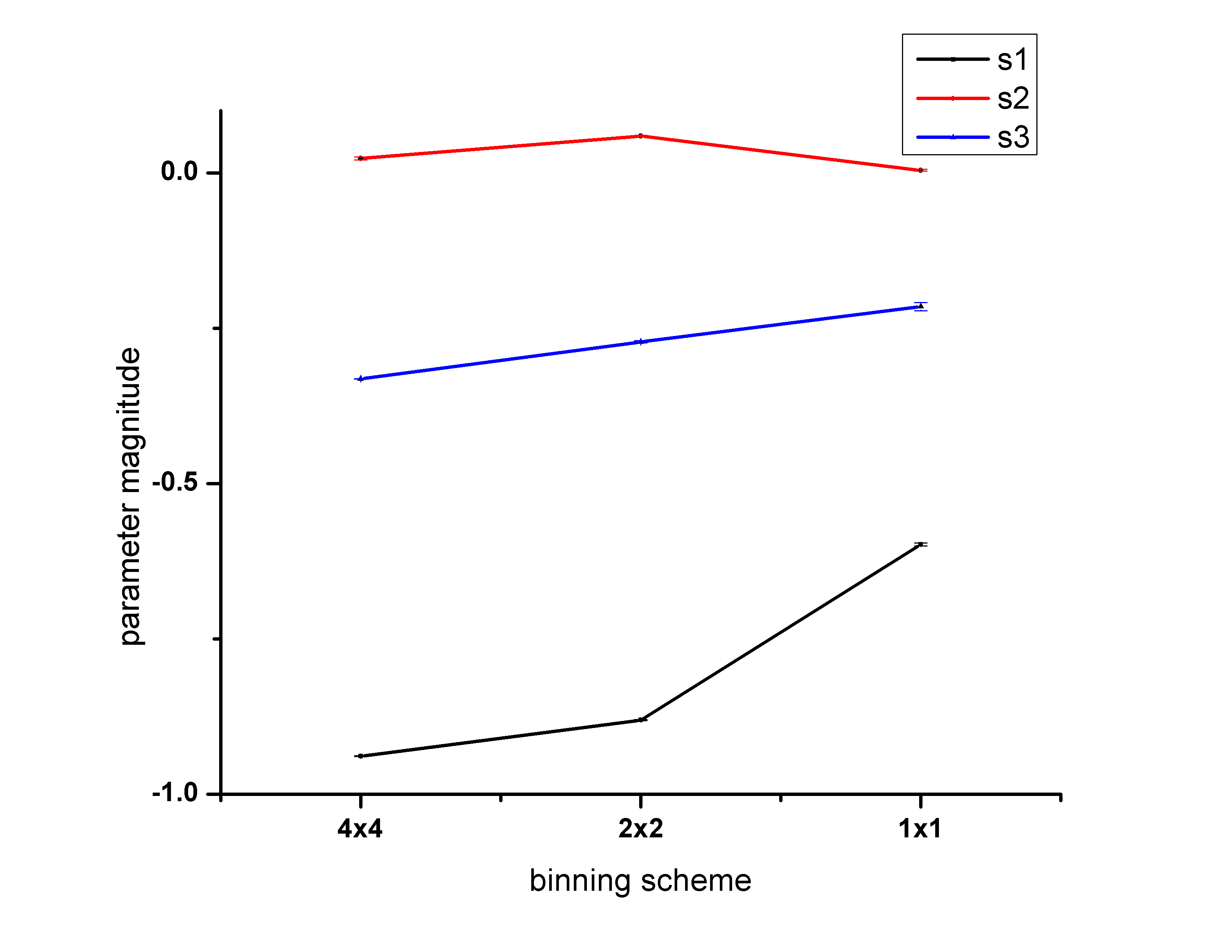}
         \caption{}
         \label{fig:a}
     \end{subfigure}
     \hfill
     \begin{subfigure}[b]{0.9\columnwidth}
         \centering
         \includegraphics[width=\textwidth]{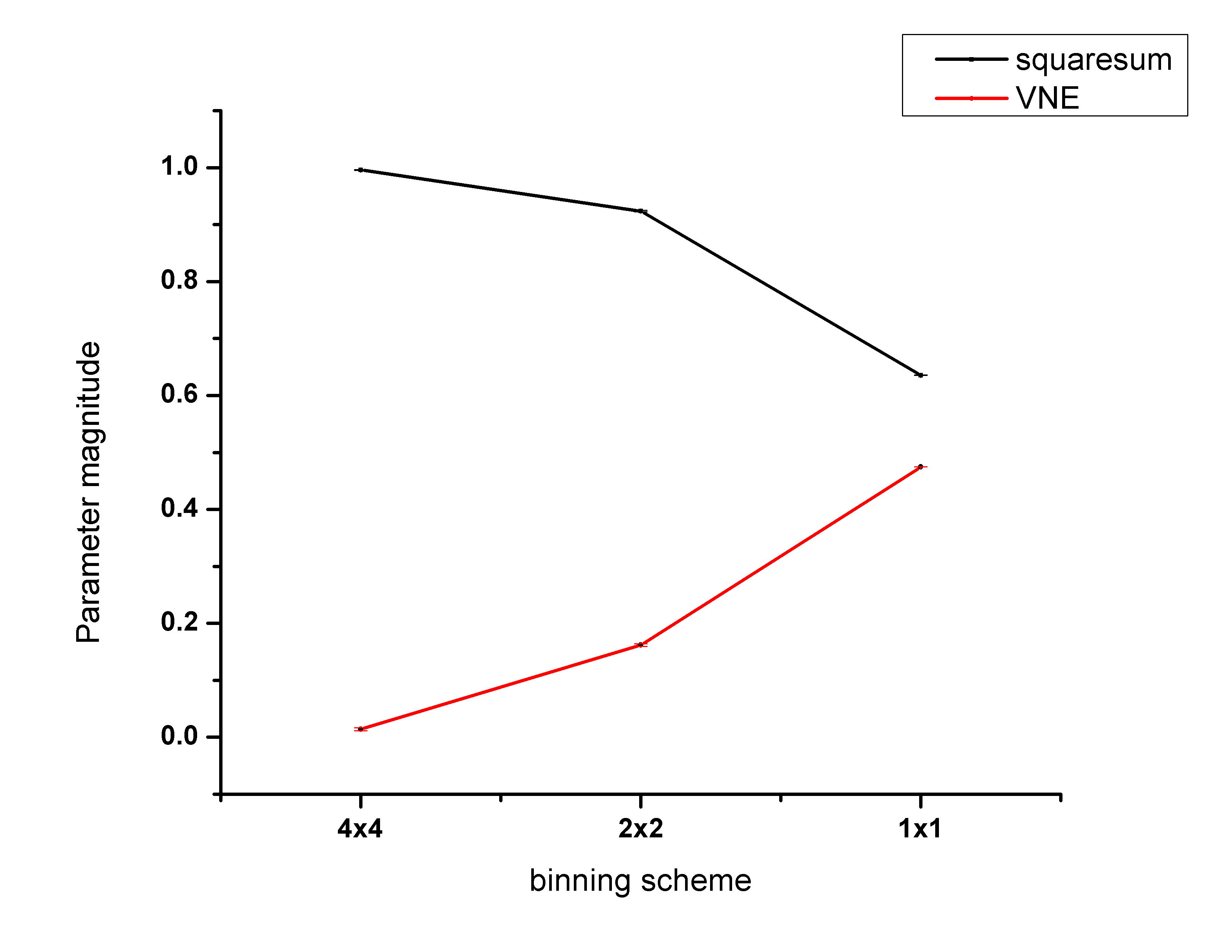}
         \caption{}
         \label{fig:b}
     \end{subfigure}
        \caption{Plots (a) The variation of each Stokes parameter with different binning settings for the diode laser; (b) The variation of VNE and squared sum across different binning conditions. }
        \label{fig:3}
\end{figure}

\FloatBarrier

\section{Light Emitting Diode}

The next source was a partially polarized LED with output peaking at 450 nm and FWHM of around 30 nm (recorded via ocean optics spectrometer, figure not shown). A linear polarizer was used to make the output partially polarized  along with a set of Neutral density filters so as to reduce the intesity to a level that is safe for the EMCCD sensor.  The light from an LED is known to be unpolarized and using the linear polarizer allowed us to get a better understanding of the effectiveness of the measurement. The basic setup remained more or less the same apart from some slight realignment.

The result of measuring the polarization parameters in this case was that while the linear polarizer introduced a certain degree of linear polarization to the light, it did not completely polarize the LED output. This shows up in the values of both square sum of the Stoke's parameters $s_1, s_2$ and $s_3$, which comes out to be a little over 0.7 and the von Neumann entropy hovers around 0.4 across the region of interest in the case of maximum binning which gives best SNR as shown in Fig. 5. This is in agreement with our calibration experiment. The effect of improved SNR is also apparent in Fig. 4, where the range of variation across the profile improves with binning.

\begin{figure}[!hbt]
     \centering
     \begin{subfigure}[b]{0.9\columnwidth}
         \centering
         \includegraphics[width=\textwidth]{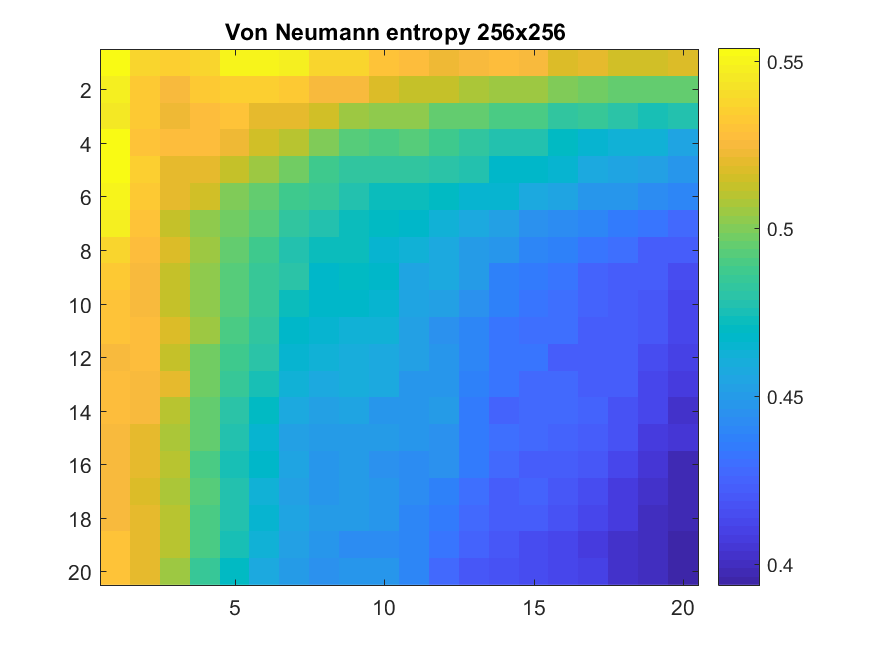}
         \caption{4$\times$4 binning}
         \label{fig:a}
     \end{subfigure}
     \hfill
     \begin{subfigure}[b]{0.9\columnwidth}
         \centering
         \includegraphics[width=\textwidth]{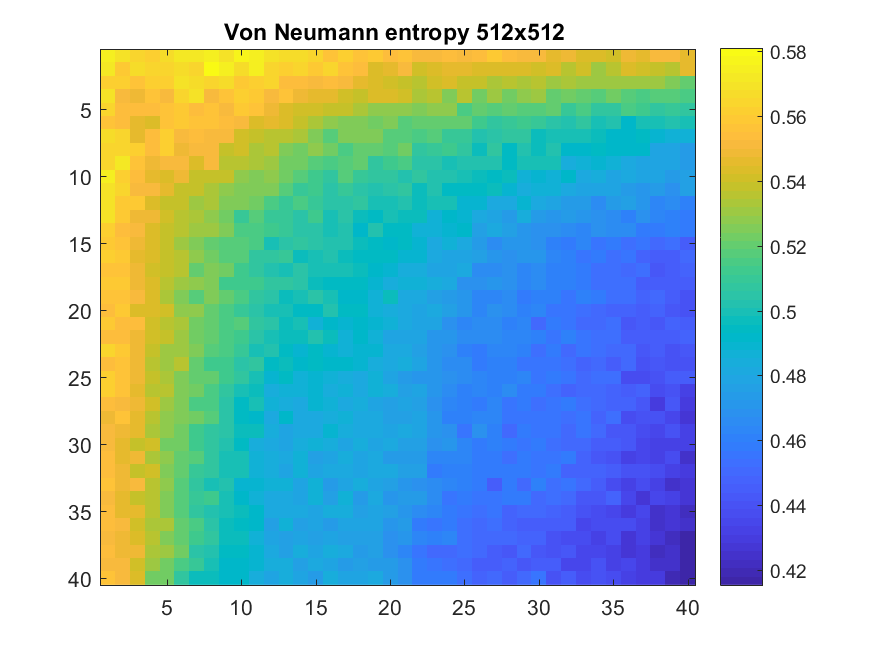}
         \caption{2$\times$2 binning}
         \label{fig:b}
     \end{subfigure}
     \hfill
     \begin{subfigure}[b]{0.9\columnwidth}
         \centering
         \includegraphics[width=\textwidth]{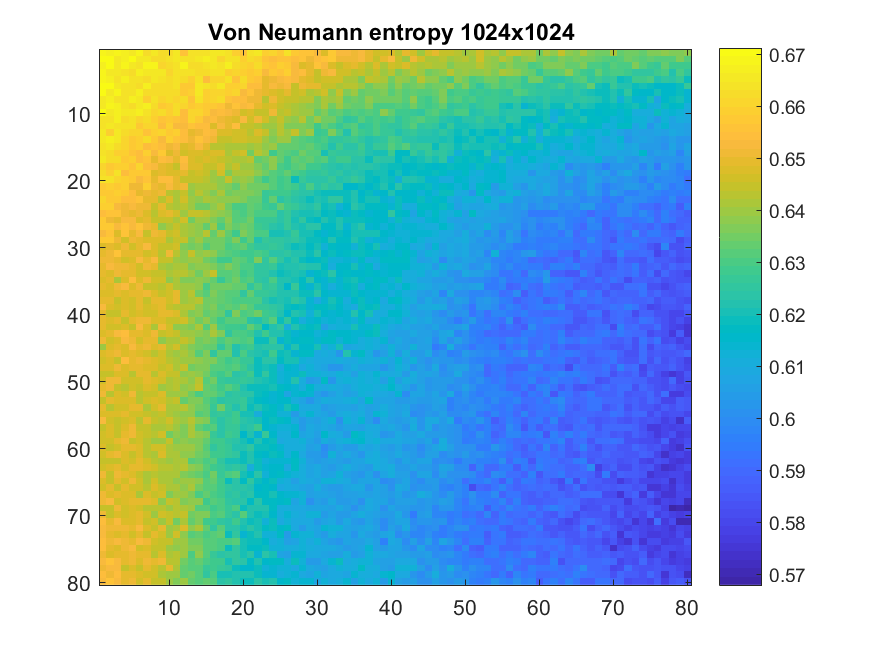}
         \caption{1$\times$1 binning}
         \label{fig:c}
     \end{subfigure}
        \caption{The spatial variation in the von Neumann entropy for different bin sizes as shown in each map title. These correspond to the region of interest in the case of LED light. The variations are once again spread over a very narrow range.}
        \label{fig:4}
\end{figure}

\begin{figure}[!hbt]
     \centering
     \begin{subfigure}[b]{0.9\columnwidth}
         \centering
         \includegraphics[width=\textwidth]{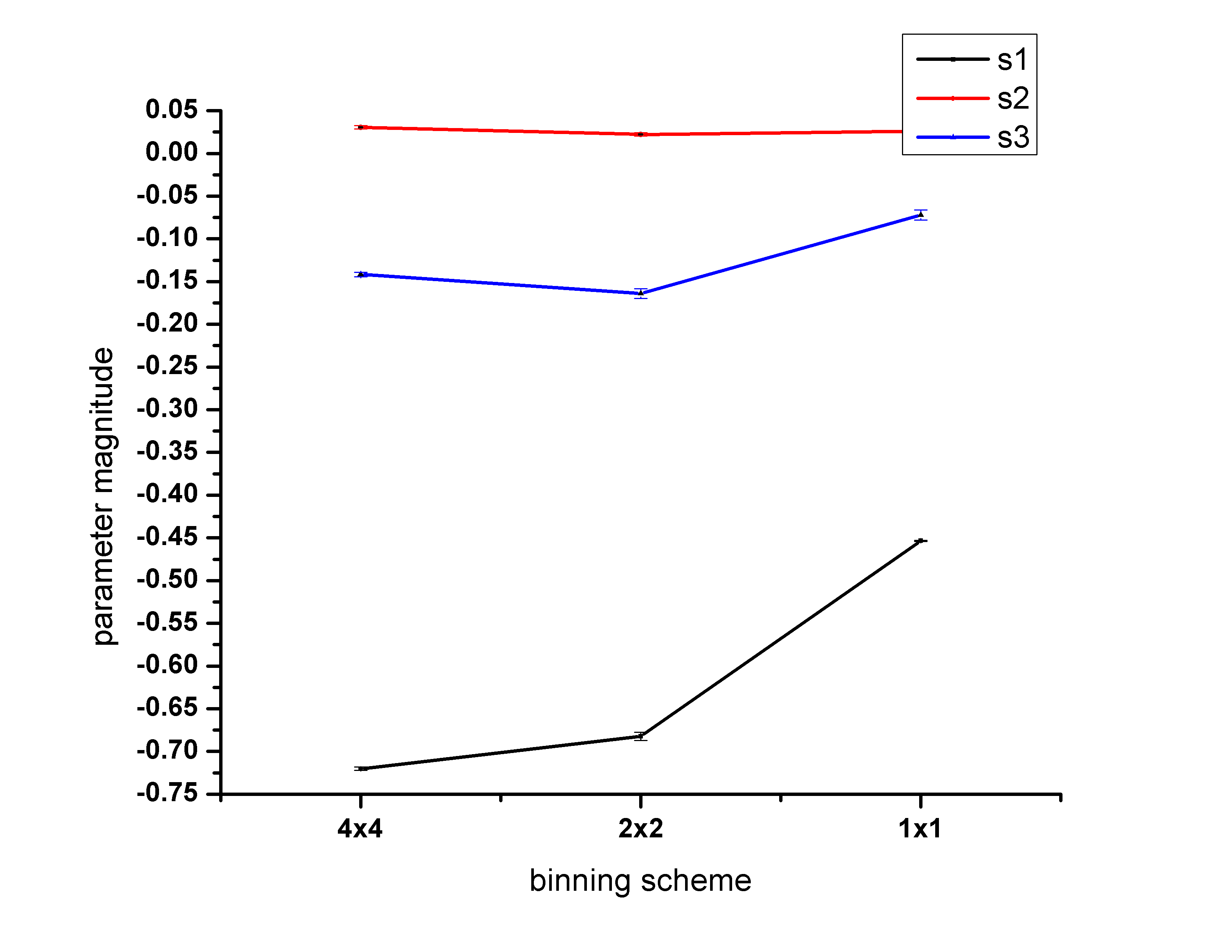}
         \caption{}
         \label{fig:a}
     \end{subfigure}
     \hfill
     \begin{subfigure}[b]{0.9\columnwidth}
         \centering
         \includegraphics[width=\textwidth]{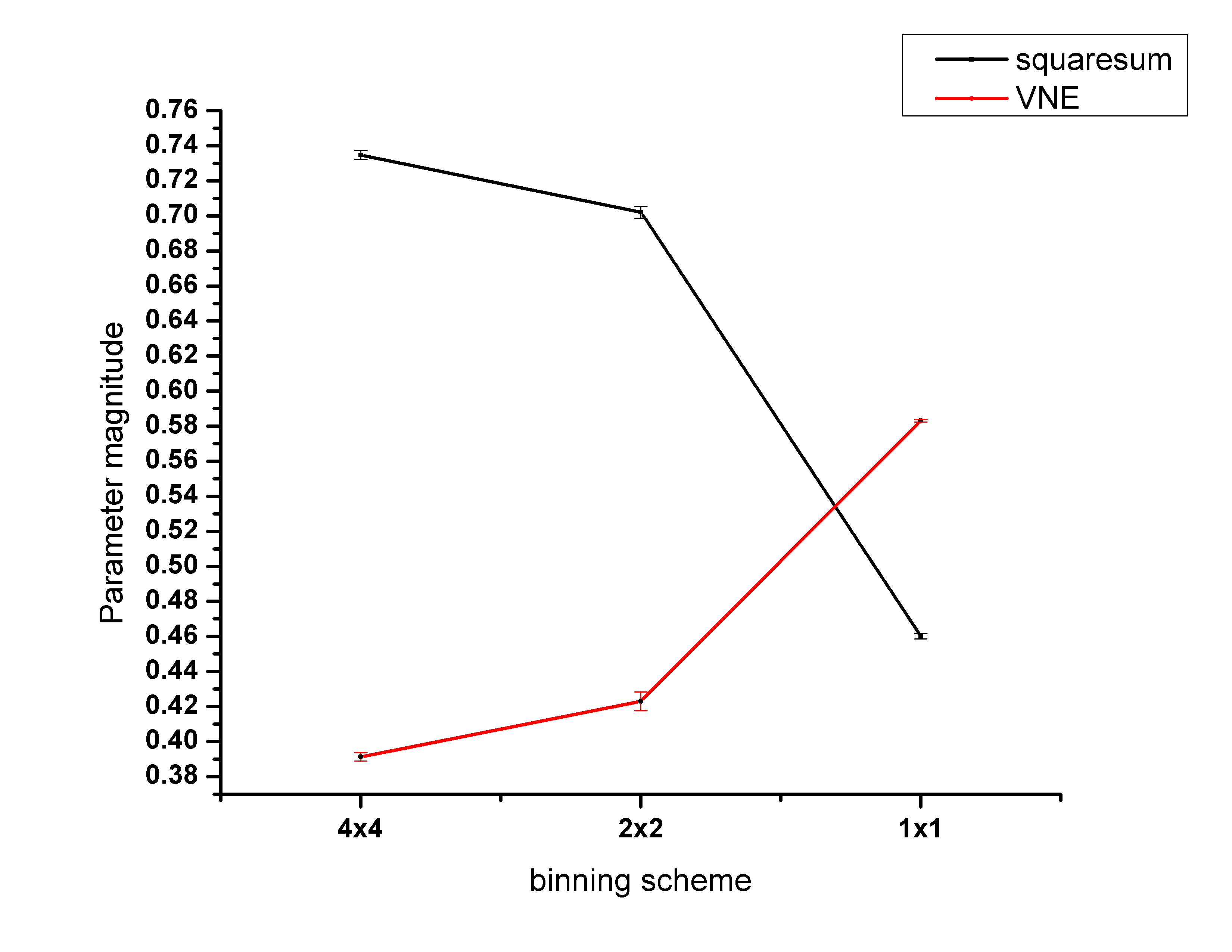}
         \caption{}
         \label{fig:b}
     \end{subfigure}
        \caption{Plots (a) The variation of each Stokes parameter with different binning settings for the LED; (b) Exhibits the inverse proportionality of the squared sum of the Stokes parameters to the von Neumann entropy for different binning conditions.}
        \label{fig:5}
\end{figure}

\FloatBarrier

\section{Fluorescein dye}

We employed a 405 nm diode laser to pump a solution of the organic fluorophore, Fluorescein which emits across a broad range from about 490 nm to 640 nm. The emission bandwidth was restricted using a 577 nm optical filter and collected and imaged using a system of lenses aligned to form an image at the EMCCD sensor. The polarimetry setup was introduced prior to the EMCCD and the Stokes measurements were recorded to study the transverse spatial variation of the fluroscent beam. Fluorescein is an organic fluorescent dye with widespread applications in bio-imaging as an indicator or tracer. Examining the transverse spatial variation of polarization entropy of this contrast agent might bring out otherwise hidden information about the sample under examination. The technique we employed can easily be modified to work with a bio-imaging setup.
As is evident from Fig. 6, the von Neumann entropy varies very little and the presence of polarization change inducing structures or processes will exhibit a noticeable change. In Fig. 7, both VNE and the square sum of the Stokes parameters appear not to change very much across different binning conditions. This can be attributed to the highly unpolarized nature of the dye emission. 

\begin{figure}[!hbt]
     \centering
     \begin{subfigure}[b]{0.9\columnwidth}
         \centering
         \includegraphics[width=\textwidth]{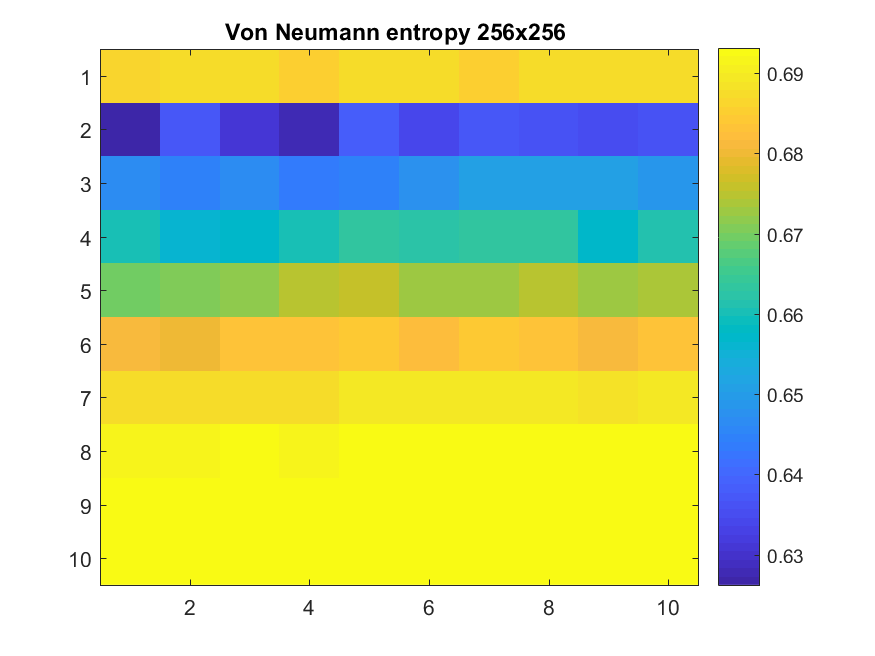}
         \caption{4$\times$4 binning}
         \label{fig:a}
     \end{subfigure}
     \hfill
     \begin{subfigure}[b]{0.9\columnwidth}
         \centering
         \includegraphics[width=\textwidth]{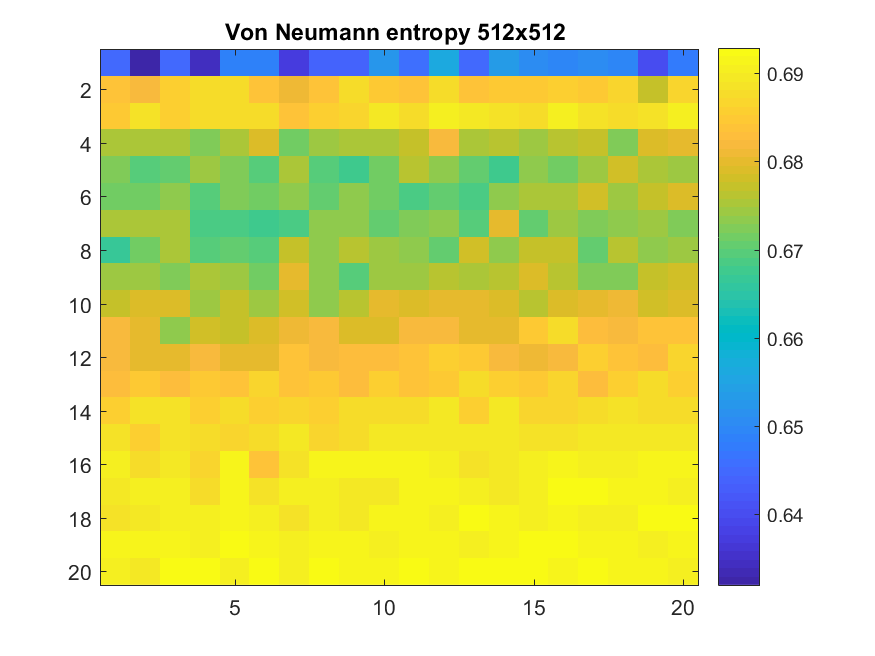}
         \caption{2$\times$2 binning}
         \label{fig:b}
     \end{subfigure}
     \hfill
     \begin{subfigure}[b]{0.9\columnwidth}
         \centering
         \includegraphics[width=\textwidth]{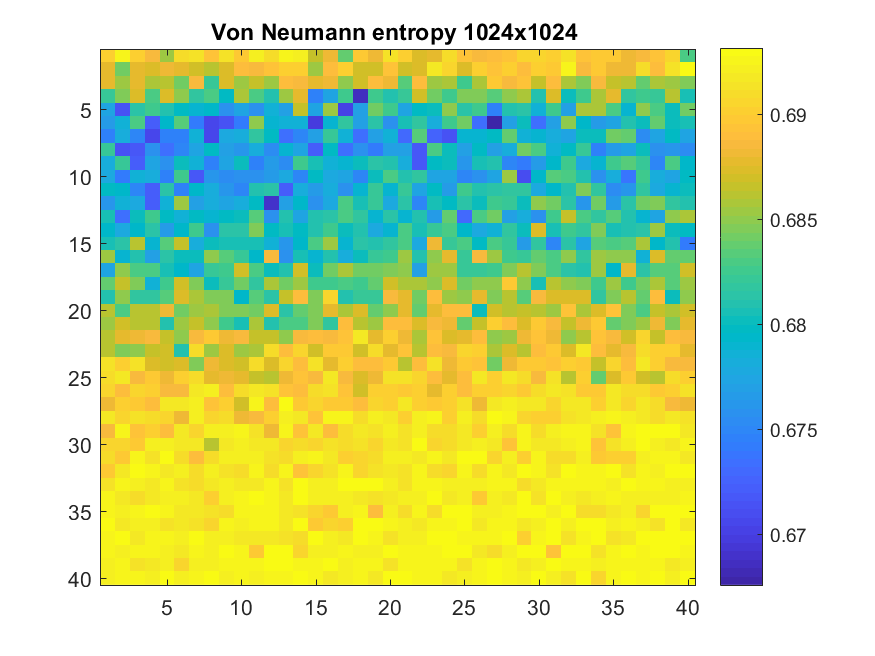}
         \caption{1$\times$1 binning}
         \label{fig:c}
     \end{subfigure}
        \caption{The spatial variation in the von Neumann entropy for different bin sizes as, shown in each map title. These correspond to the region of interest in the case of fluorophore emission output.}
        \label{fig:6}
\end{figure}

\begin{figure}[!hbt]
     \centering
     \begin{subfigure}[b]{0.9\columnwidth}
         \centering
         \includegraphics[width=\textwidth]{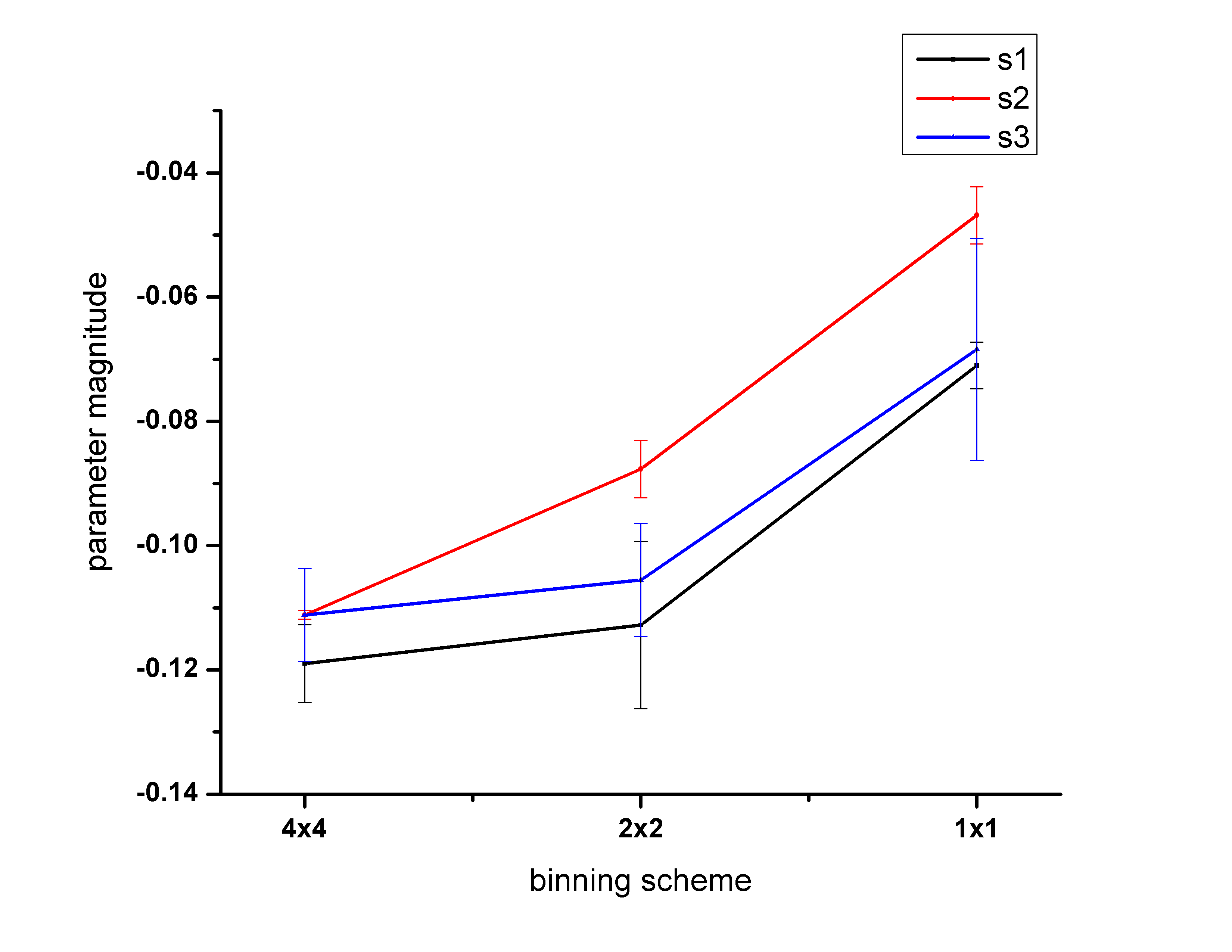}
         \caption{}
         \label{fig:a}
     \end{subfigure}
     \hfill
     \begin{subfigure}[b]{0.9\columnwidth}
         \centering
         \includegraphics[width=\textwidth]{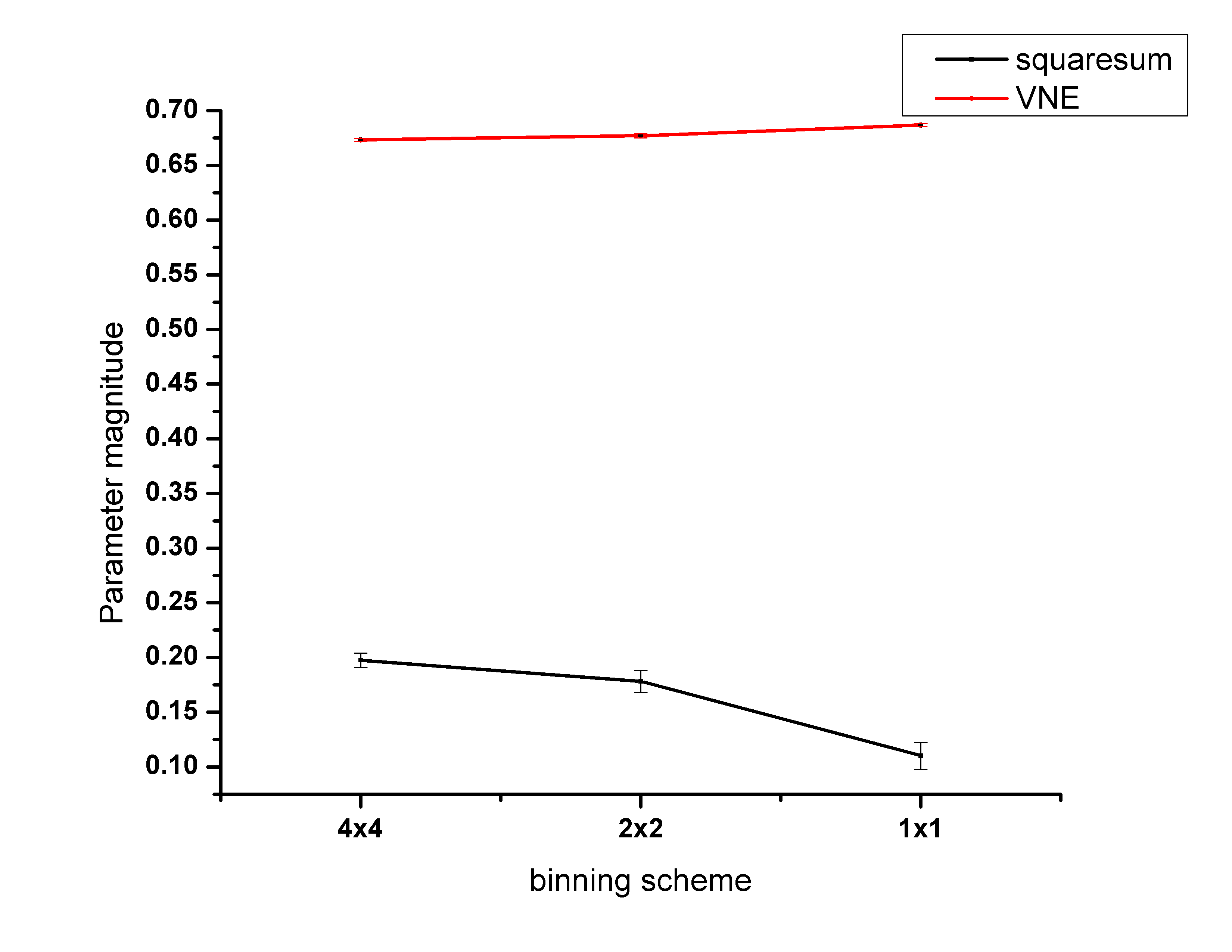}
         \caption{}
         \label{fig:b}
     \end{subfigure}
        \caption{Plots (a) The variation of each Stokes parameter with different binning settings for the fluorescent dye; (b) Exhibits the inverse proportionality of the squared sum of the Stokes parameters to the von Neumann entropy for different binning conditions.}
        \label{fig:7}
\end{figure}

\FloatBarrier

\section{Spontaneous Parametric Down Conversion}

Type-II Spontaneous parametric down conversion (SPDC) [17] output has a cross section in the shape of twin rings intersecting at two points, as shown in Fig. 10. While the rings are orthogonally polarized w.r.t each other, their points of intersection at any given instant can have either vertical or horizontal polarization. At these points of intersection, the polarizations across these points will always be anti-correlated but it is impossible to predict the polarization at any one point at any given instant. We wanted to see if this randomness would reflect in the von Neumann entropy measurement of these points of intersection.

To measure the polarization properties at the points of intersection, we first had to image the twin rings of SPDC without any of the optics associated with polarimetric in the way. Once SPDC was confirmed, a telescope type optical system was introduced between the EMCCD and nonlinear crystal to allow for space in between to introduce the polarimetry specific optics. Custom apertures were used to isolate the points of intersection of the twin rings in the image.
 Once again, the variations across our sensor area are minimal but become slighly more pronounced at higher binning stages, as shown in Fig. 8. In Fig. 9, the variation of the squared sum of the Stokes parameters is more drastic than that of VNE, hinting at an inverse but nonlinear relationship between the two parameters.
 
\begin{figure}[!hbt]
     \centering
     \begin{subfigure}[b]{0.9\columnwidth}
         \centering
         \includegraphics[width=\textwidth]{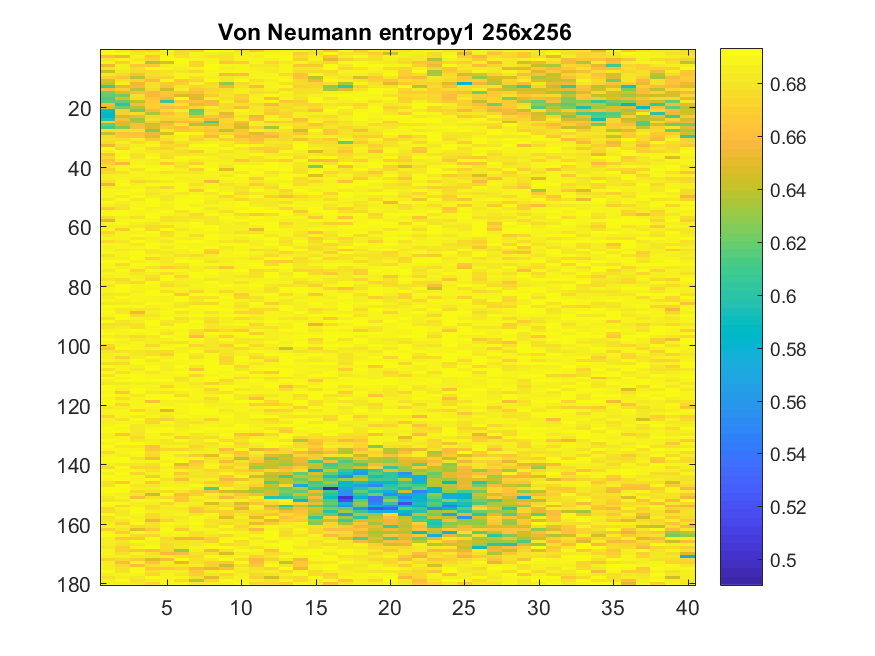}
         \caption{4$\times$4 binning}
         \label{fig:a}
     \end{subfigure}
     \hfill
     \begin{subfigure}[b]{0.9\columnwidth}
         \centering
         \includegraphics[width=\textwidth]{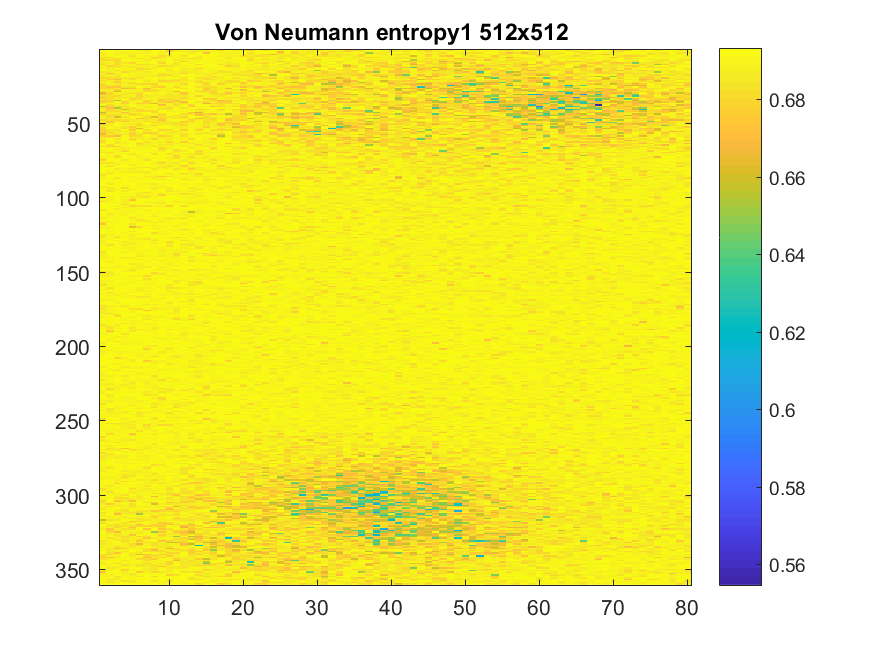}
         \caption{2$\times$2 binning}
         \label{fig:b}
     \end{subfigure}
     \hfill
     \begin{subfigure}[b]{0.9\columnwidth}
         \centering
         \includegraphics[width=\textwidth]{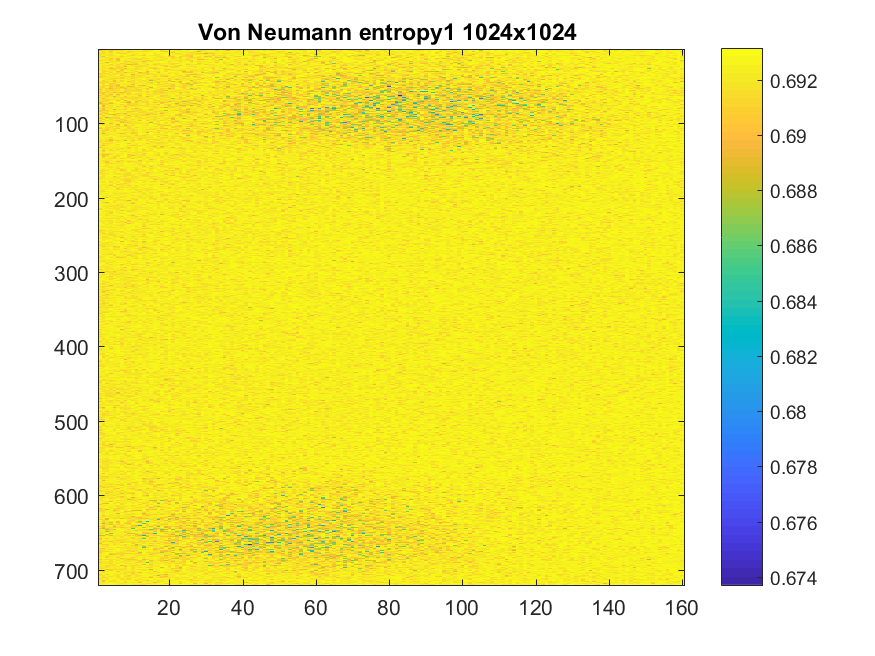}
         \caption{1$\times$1 binning}
         \label{fig:c}
     \end{subfigure}
        \caption{The spatial variation in the von Neumann entropy for different bin sizes as shown in each map title. These correspond to the region of interest in the case of SPDC output.}
        \label{fig:8}
\end{figure}

\begin{figure}[!hbt]
     \centering
     \begin{subfigure}[b]{0.9\columnwidth}
         \centering
         \includegraphics[width=\textwidth]{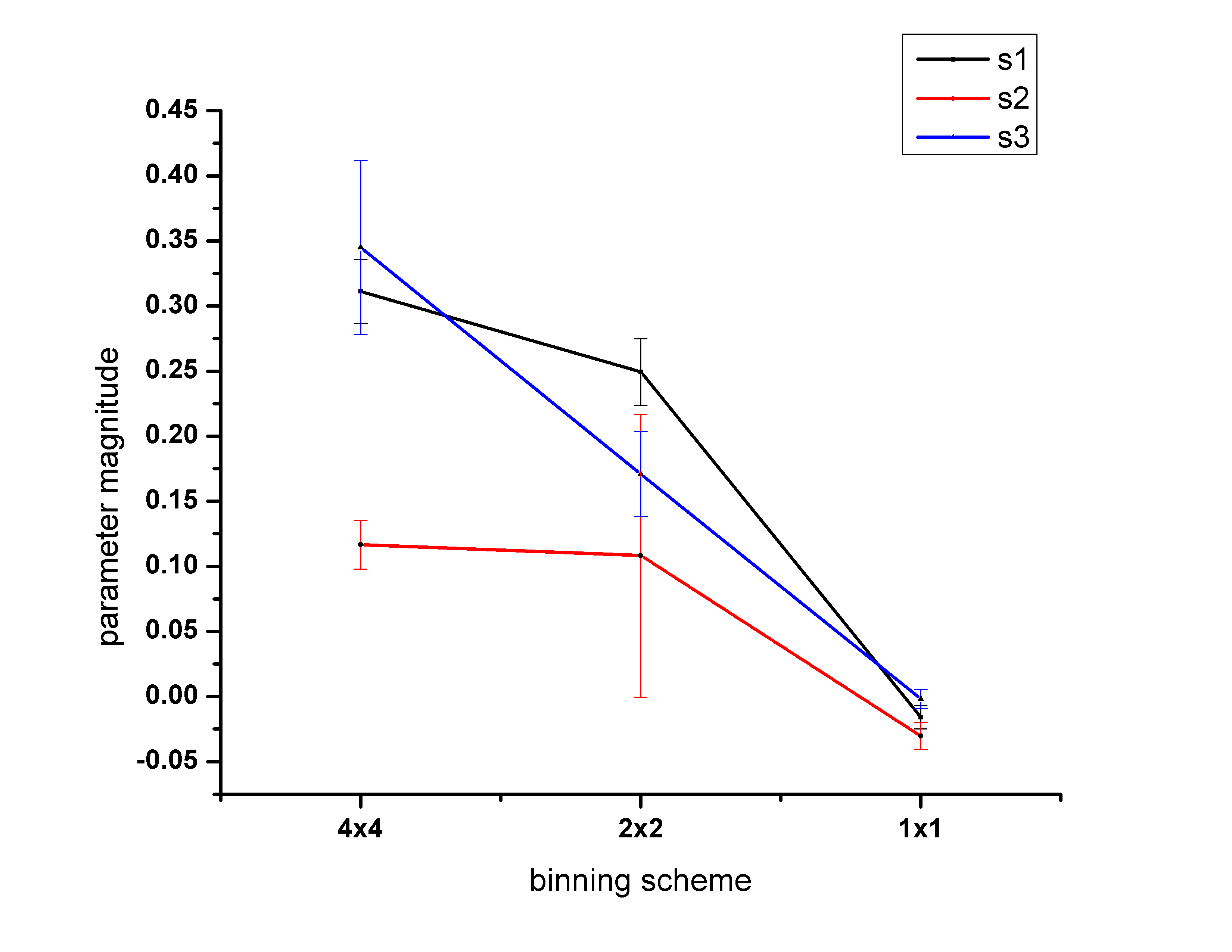}
         \caption{}
         \label{fig:a}
     \end{subfigure}
     \hfill
     \begin{subfigure}[b]{0.9\columnwidth}
         \centering
         \includegraphics[width=\textwidth]{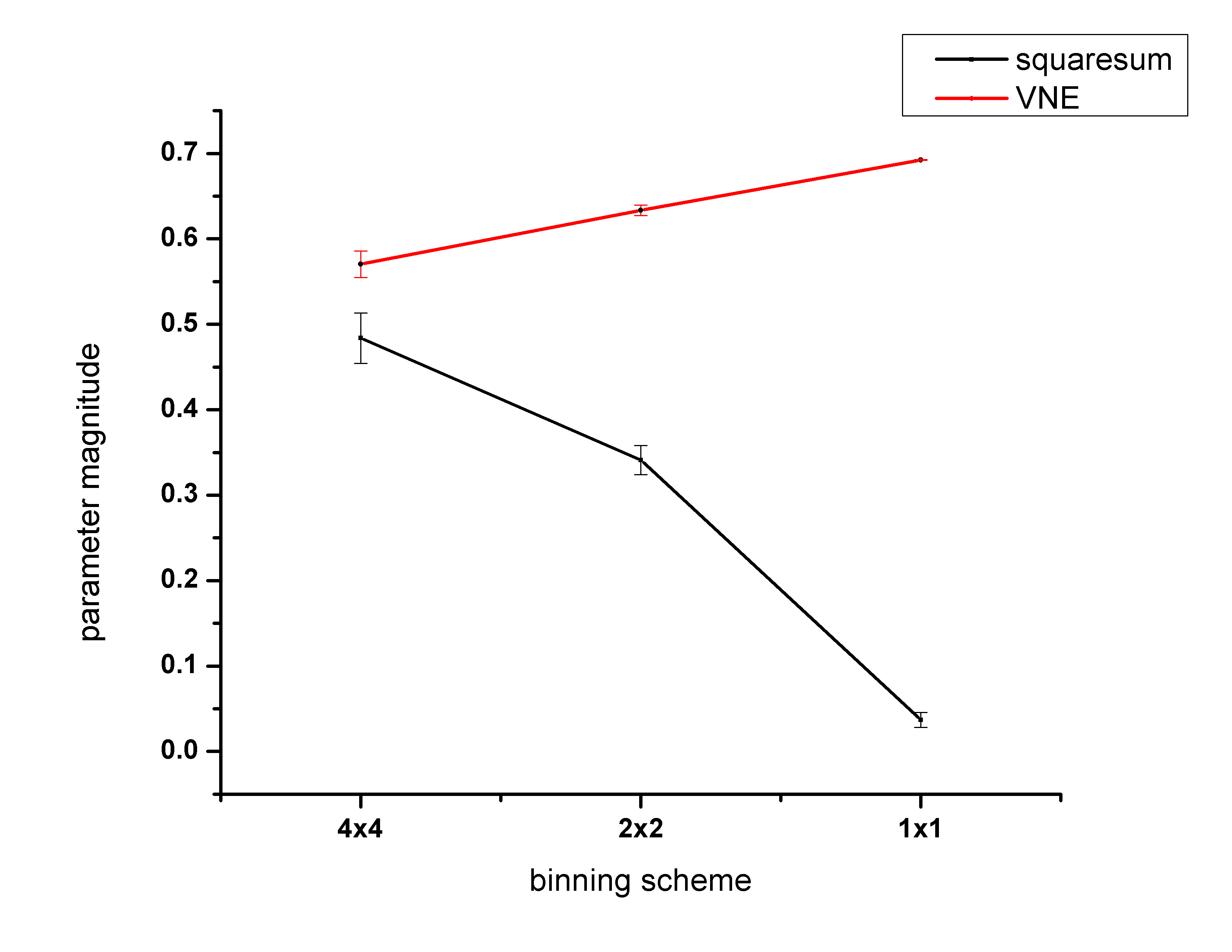}
         \caption{}
         \label{fig:b}
     \end{subfigure}
        \caption{(a) The variations of the stokes parameters and (b) The variation of VNE and square sum across different stages of binning in the case of SPDC.}
        \label{fig:9}
\end{figure}

\begin{figure}[!hbt]
\centering     
\includegraphics[width=70mm]{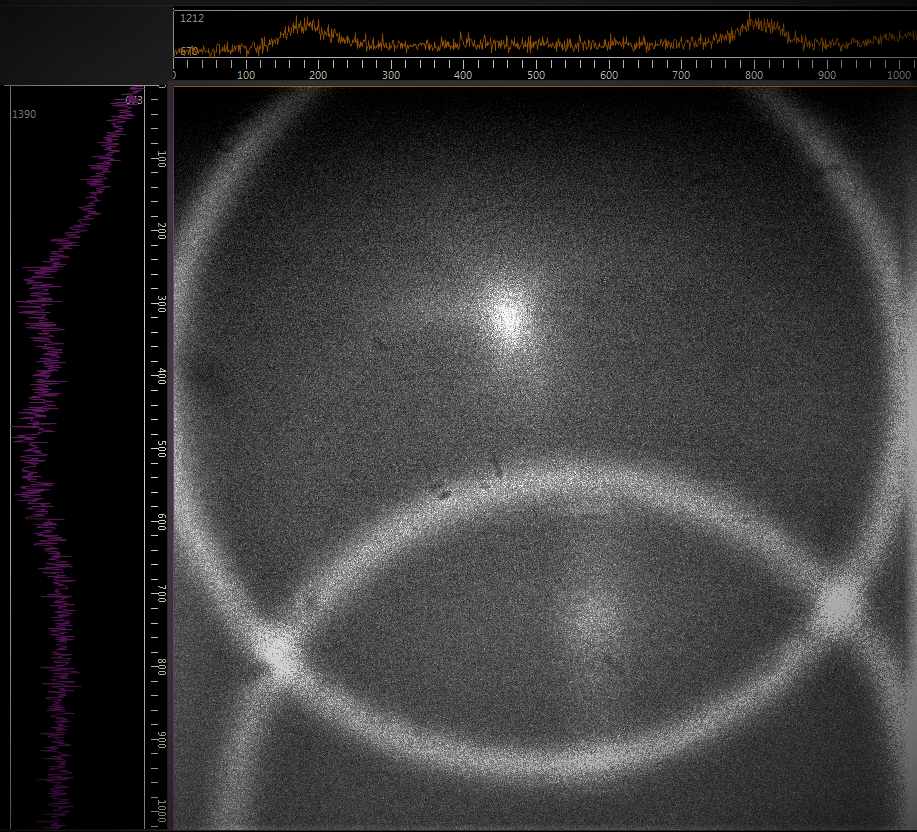}
\caption{The image of SPDC rings recorded using EMCCD.}
\label{fig:10}
\end{figure}

\FloatBarrier

\section{Results and Discussion}
We have studied the polarization properties across the beam profile of several light sources. The highly polarized laser output was found to have the least von Neumann entropy associated with polarization, while the emission from fluorescein dye, along with SPDC output showed the highest values. However, the von Neumann entropy values of a few small regions within the laser beam spot appear to take on very small negative values while the majority of of the area shows values just above zero. While we expect the small positive VNE values for the highly polarized output of the diode laser, the slight negative deviations can be attributed to experimentally induced effects.

The effect of binning on the VNE of the different sources, as shown in Fig. 11(b), suggests that more accurate values are obtained with higher binning; the well defined linear polarization of the diode laser can be thought to have less disorder or entropy and the lowest value of VNE is obtained at the highest possible stage of binning. A similar argument can be made for the partially polarized output from the LED. The emission of the Fluorescein dye, being somewhat unpolarized, showed not much variation in its VNE values across different stages of binning and also the least amount of variation within the region of interest. The correlation between degree of polarization and VNE is evident in Fig. 11(a) where the unpolarized emission of the dye shows the highest value of VNE of all the sources. The VNE of SPDC is also high. This is because where the dual rings of SPDC intersect, the polarization toggles randomly between horizontal and vertical and our measurements show that this gives it a fairly high VNE value which also varies more across the region, as compared to that of the dye.

The range of variation of the Stokes parameter $s_{1}$, across the region of interest for the different sources shows the expected results with the laser values varying across a very narrow range centred close to -1, which is as expected for the state of polarization. The standout feature in the plot is the large range of variation of SPDC output derived $s_{1}$ centred at around zero which we attributed to the fast toggling between the horizontal and vertical states of polarization at the points of intersection.

\begin{figure}
     \centering
     \begin{subfigure}[b]{0.9\columnwidth}
         \centering
         \includegraphics[width=0.8\columnwidth, height = 45mm]{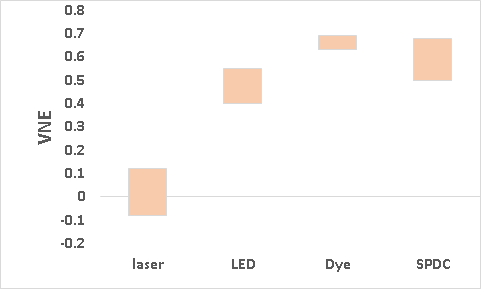}
         \caption{}
         \label{fig:a}
     \end{subfigure}
     \hfill
     \begin{subfigure}[b]{0.9\columnwidth}
         \centering
         \includegraphics[width=0.8\columnwidth, height = 45mm]{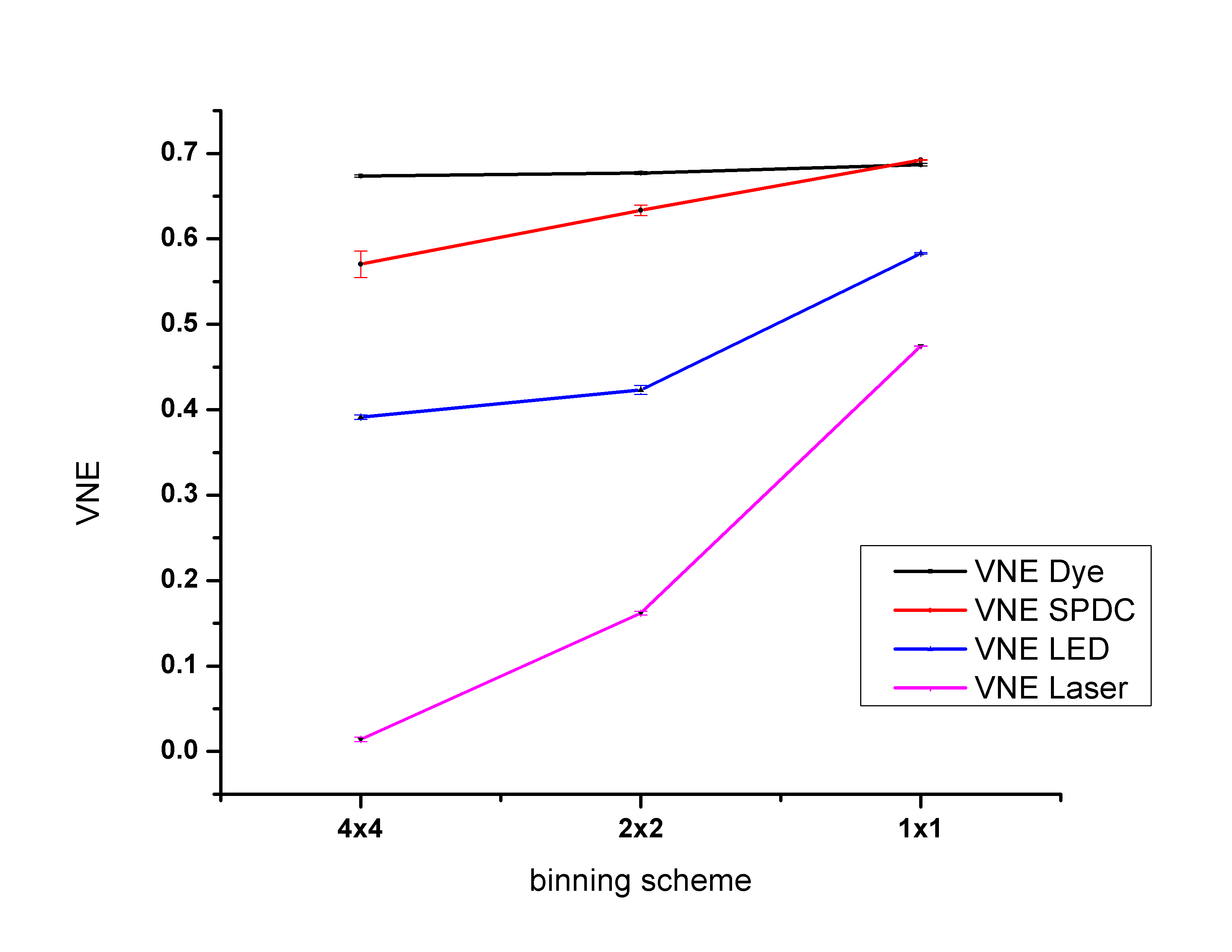}
         \caption{}
         \label{fig:b}
     \end{subfigure}
     \begin{subfigure}[b]{0.9\columnwidth}
         \centering
         \includegraphics[width=0.8\columnwidth, height = 45mm]{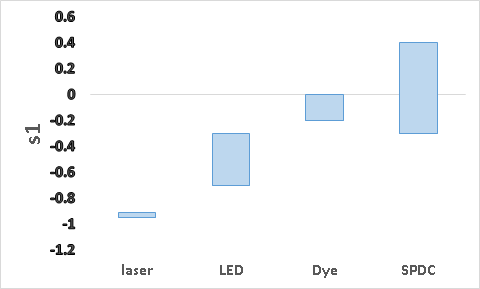}
         \caption{}
     \end{subfigure}
     \hfill
     \begin{subfigure}[b]{0.9\columnwidth}
         \centering
         \includegraphics[width=0.8\columnwidth, height = 45mm]{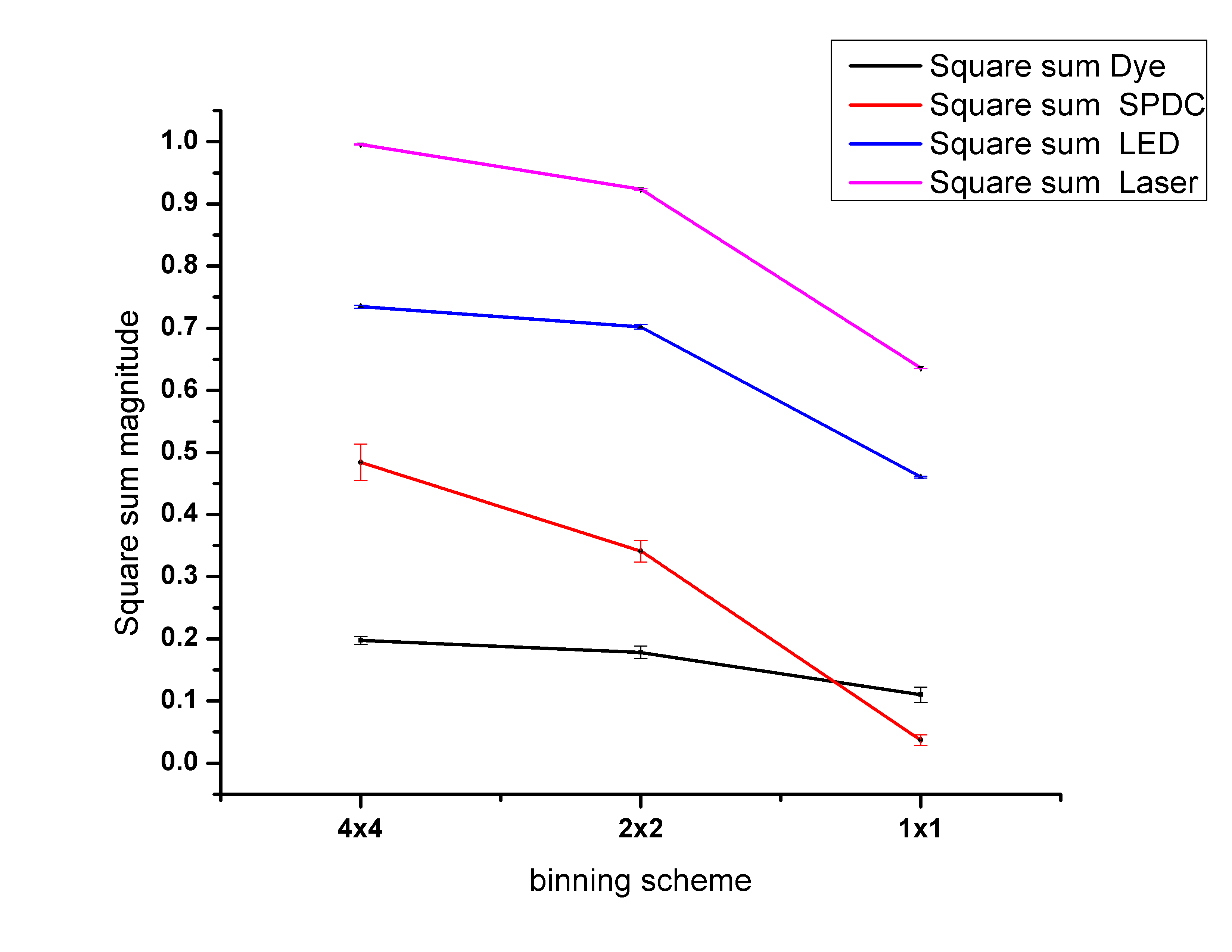}
         \caption{}
     \end{subfigure}
        \caption{(a) The range of variation of VNE within the region of interest for the different sources, (b) A comparison between different sources in terms of how the VNE varies across different stages of binning, (c) The range of variation of $s_{1}$ within the region of interest for all the investigated sources and (d) The variation of square sum of Stokes parameters entropy across different binning conditions for all the investigated sources.}
        \label{fig:11}
\end{figure}

\FloatBarrier

\section{Conclusions}

To conclude, we have measured the Stokes parameters as well as von Neumann entropy and their transverse spatial variation across the illuminated area using our simple setup for the light from a CW diode laser, an LED, a fluorescent dye and the entangled output of SPDC. Furthermore, we studied the effects of binning on all measured parameters which we found to be in step with the improvement in SNR or visibility that the process of binning brings about. There is a clear trend of increasing von Neumann entropy with decreasing polarization purity which is as expected but there is also a decrease of the von Neumann entropy with degree of binning.  These results can be incorporated positively in the field of Synthetic Aperture Radar polarimetry which employs polarization entropy concepts to map and lock on to targets spread out over a large area.

\section*{Acknowledgments}
VN and SB  acknowledge  the support from Interdisciplinary Cyber Physical Systems (ICPS) programme of the Department of Science and Technology (DST), India, Grant No.:DST/ICPS/QuST/Theme-1/2019/6.
Tomis would like to acknowledge CSIR for the research funding. RP acknowledges DST for the finanancial assitance given for the research.

\bigskip

\end{document}